\documentclass[12pt,titlepage]{article}
\usepackage{amssymb}
\usepackage[dvips]{graphicx}
\usepackage{amsfonts}
\usepackage[portuguese]{babel}

\begin{document}

\title{\textbf{Estados estacion\'{a}rios de part\'{\i}culas sem spin em
potenciais quadrados}\\
{\small \textbf{Stationary states of spinless particles in square potentials}%
}}
\author{Tatiana R. Cardoso e Antonio S. de Castro\thanks{%
E-mail: castro@pesquisador.cnpq.br.} \\
\\
Departamento de F\'{\i}sica e Qu\'{\i}mica\\
Universidade Estadual Paulista\\
Guaratinguet\'{a} SP - Brasil}
\date{}
\maketitle

\begin{abstract}
\noindent A equa\c{c}\~{a}o de Klein-Gordon em uma dimens\~{a}o espacial
\'{e} investigada com a mais geral estrutura de Lorentz para os potenciais
externos. A an\'{a}lise e o c\'{a}lculo dos coeficientes de reflex\~{a}o e
transmiss\~{a}o para o espalhamento de part\'{\i}culas em um potencial
quadrado, com uma mistura arbitr\'{a}ria de acoplamentos vetorial e escalar,
revelam circunst\^{a}ncias que conflitam com as previs\~{o}es da mec\^{a}%
nica qu\^{a}ntica n\~{a}o-relativ\'{\i}stica. Mostra-se que tais
es\-pa\-lha\-men\-tos an\^{o}malos s\~{a}o mediados por estados ligados de
antipart\'{\i}culas, ainda que as part\'{\i}culas incidentes tenham baixas
energias. A an\'{a}lise dos estados ligados tamb\'{e}m revela resultados
surpreendentes, destacando-se a inibi\c{c}\~{a}o do efeito
Schiff-Snyder-Weinberg pela presen\c{c}a de um acoplamento escalar.\newline
\newline
\noindent \textbf{Palavras-chave:} equa\c{c}\~{a}o de Klein-Gordon, part%
\'{\i}culas sem spin, paradoxo de Klein, produ\c{c}\~{a}o de pares, efeito
Schiff-Snyder-Weinberg \newline
\newline
\newline
\noindent {\small {The one-dimensional Klein-Gordon equation is investigated with the
most general Lorentz structure for the external potentials. The analysis and
calculation of the reflection and transmission coefficients for the
scattering of particles in a square potential, with an arbitrary mixing of
vector and scalar couplings, reveal circumstances which conflict with the
predictions from nonrelativistic quantum mechanics. It is shown that such
anomalous scatterings are mediated by antiparticle bound states, even if the
incident particles have low energies. The analysis of bound states also
reveals surprising results, remarkable is the inhibition of the
Schiff-Snyder-Weinberg effect due to the presence of a scalar coupling.}
}
\end{abstract}

\section{Introdu\c{c}\~{a}o}

A generaliza\c{c}\~{a}o da mec\^{a}nica qu\^{a}ntica que inclui a
relatividade especial \'{e} ne\-ces\-s\'{a}\-ria para a descri\c{c}\~{a}o de
fen\^{o}menos em altas energias e tamb\'{e}m para a descri\c{c}\~{a}o de fen%
\^{o}menos em escalas de comprimentos que s\~{a}o menores ou compar\'{a}veis
com o comprimento de onda Compton da part\'{\i}cula ($\lambda =\hbar /(mc)$%
). A generaliza\c{c}\~{a}o n\~{a}o \'{e} uma tarefa trivial e novos e
peculiares fen\^{o}menos surgem na Mec\^{a}nica Qu\^{a}ntica Relativ\'{\i}%
stica (doravante denominada MQR). Entre tais fen\^{o}menos est\~{a}o a produ%
\c{c}\~{a}o espont\^{a}nea de pares mat\'{e}ria-antimat\'{e}ria e a limita%
\c{c}\~{a}o para a lo\-ca\-li\-za\-\c{c}\~{a}o de part\'{\i}culas. Essa
limita\c{c}\~{a}o pode ser estimada \ pela observa\c{c}\~{a}o que a m\'{a}%
xima incerteza para o momento da part\'{\i}cula $\Delta p=mc$ conduz, via
princ\'{\i}pio da incerteza de Heisenberg, \`{a} incerteza m\'{\i}nima na
posi\c{c}\~{a}o $\Delta x=\lambda /2$ \cite{gre}-\cite{str}. Embora a MQR
como modelo de part\'{\i}cula \'{u}nica, referida como formalismo de
primeira quantiza\c{c}\~{a}o, n\~{a}o possa dar conta da completa descri\c{c}%
\~{a}o da cria\c{c}\~{a}o de pares, ela pavimenta o caminho para o
desenvolvimento da Teoria Qu\^{a}ntica de Campos.

As mais simples equa\c{c}\~{o}es da MQR s\~{a}o a equa\c{c}\~{a}o de
Klein-Gordon (EKG)\footnote{%
A EKG descreve o comportamento de b\'{o}sons de spin 0. P\'{\i}ons e k\'{a}%
ons, por exemplo.} e a equa\c{c}\~{a}o de Dirac\footnote{%
A equa\c{c}\~{a}o de Dirac descreve o comportamento de f\'{e}rmions de spin
1/2, tais como o el\'{e}tron, o neutrino, o quark, o pr\'{o}ton e o n\^{e}%
utron.
\par
{}}. O spin \'{e} uma complica\c{c}\~{a}o adicional na MQR e,
na\-tu\-ral\-men\-te, a EKG permite que certos aspectos da MQR possam ser
a\-na\-li\-sa\-dos com um formalismo matem\'{a}tico mais simples e
percebidos com maior transpar\^{e}ncia.

A solu\c{c}\~{a}o da equa\c{c}\~{a}o de Dirac para o espalhamento de part%
\'{\i}culas em um potencial degrau, considerado como o componente temporal
de um potencial vetorial, \'{e} bem conhecida e cristalizada em livros-texto
\cite{gre}-\cite{gro}. Neste pro\-ble\-ma surge o c\'{e}lebre paradoxo de
Klein \cite{kle} para potenciais suficientemente intensos, um fen\^{o}meno
em que o coeficiente de reflex\~{a}o excede a unidade e \'{e} interpretado
como sendo devido \`{a} cria\c{c}\~{a}o de pares na interface do potencial.
A an\'{a}lise do problema consoante a EKG n\~{a}o foi esquecida \cite{gro},
\cite{wi}-\cite{tati}.

O espalhamento relativ\'{\i}stico de part\'{\i}culas por barreiras de
potenciais quadradas com acoplamento puramente vetorial tem sido estudado
com a equa\c{c}\~{a}o de Dirac \cite{gre},\cite{tho}-\cite{dom} e tamb\'{e}m
com a equa\c{c}\~{a}o de KG \cite{gre}. Em ambos os casos, observa-se
transmiss\~{a}o ressonante para barreiras suficientemente altas, quando ent%
\~{a}o a mec\^{a}nica qu\^{a}ntica \ n\~{a}o-relativ\'{\i}stica prev\^{e} a
supress\~{a}o exponencial do coeficiente de transmiss\~{a}o. Tal transmiss%
\~{a}o ressonante relat\'{\i}vistica \'{e} associada com o processo de cria%
\c{c}\~{a}o de pares part\'{\i}cula-antipart\'{\i}cula na regi\~{a}o de
intera\c{c}\~{a}o. Os estados ligados da equa\c{c}\~{a}o de KG, por\'{e}m,
diferem radicalmente daqueles da equa\c{c}\~{a}o de Dirac. Para po\c{c}os de
potenciais vetoriais estreitos e profundos o bastante, a equa\c{c}\~{a}o de
KG ostenta o fen\^{o}meno conhecido na literatura como efeito
Schiff-Snyder-Weinberg (SSW) \cite{ssw}. Tal efeito manifesta-se pelo
surgimento de estados ligados de antipart\'{\i}culas adicionais num
potencial que \'{e} atrativo somente para part\'{\i}culas\footnote{%
Um potencial vetorial acopla com a carga el\'{e}trica e se ele for atrativo
(repulsivo) para cargas positivas ser\'{a} repulsivo (atrativo) para cargas
negativas.}. Para profundidades cr\'{\i}ticas, os n\'{\i}veis de energia dos
estados ligados de par\-t\'{\i}\-cu\-las e antipart\'{\i}culas coalescem e a%
\'{\i} se descortina um novo canal para a produ\c{c}\~{a}o espont\^{a}nea de
pares part\'{\i}cula-antipart\'{\i}cula. Popov \cite{pop} advogou que o
efeito SSW \'{e} caracter\'{\i}stico de potenciais de curto alcance e que n%
\~{a}o de\-ve\-ria ser esperado para potenciais de longo alcance. Contudo,
Klein e Rafelski \cite{kle2} usaram um suposto efeito SSW em um po\c{c}o
coulombiano para es\-pe\-cu\-lar sobre a condensa\c{c}\~{a}o de Bose e a
estabilidade de n\'{u}cleos com n\'{u}meros at\^{o}micos extremamente altos
e, de imediato, foram severamente criticados \cite{baw}. De fato, a investiga%
\c{c}\~{a}o dos estados ligados da equa\c{c}\~{a}o de KG com
di\-fe\-ren\-tes formas funcionais para os potenciais vetoriais confirmam a
suposi\c{c}\~{a}o do Popov \cite{flei}-\cite{villalba}.

Na Ref. \cite{tati} analisamos a EKG unidimensional com intera\c{c}\~{o}es
externas com a mais geral estrutura de Lorentz, i.e., consideramos
potenciais com estrutura vetorial, com componentes espacial e temporal,
acrescido de uma estrutura escalar. Em seguida exploramos as solu\c{c}\~{o}%
es para o espalhamento de part\'{\i}culas em um potencial degrau com
acoplamento geral, por assim dizer, com uma mistura arbitr\'{a}ria de
acoplamentos vetorial e escalar. Verificamos que tal mistura de acoplamentos
conduz a resultados surpreendentes. Para al\'{e}m de aumentar o limiar de
energia para a produ\c{c}\~{a}o espont\^{a}nea de pares, podendo at\'{e}
mesmo frustrar a produ\c{c}\~{a}o ainda que os potenciais sejam extremamente
fortes, a presen\c{c}a de um acoplamento escalar permite que uma part\'{\i}%
cula possa ser localizada em uma regi\~{a}o do espa\c{c}o arbitrariamente
pequena sem amea\c{c}ar a interpreta\c{c}\~{a}o de part\'{\i}cula \'{u}nica
da EKG. A aparente viola\c{c}\~{a}o do princ\'{\i}pio da incerteza foi
remediada com a introdu\c{c}\~{a}o do conceito de comprimento de onda
Compton efetivo.

O prop\'{o}sito do presente trabalho \'{e} dar seguimento ao estudo da EKG
unidimensional levado a cabo na Ref. [12]. Desta feita analisamos os estados
estacion\'{a}rios em um potencial quadrado com uma mistura arbitr\'{a}ria de
acoplamentos vetorial e escalar. Calculamos os coeficientes de reflex\~{a}o
e transmiss\~{a}o e obtemos resultados que contradizem a intui\c{c}\~{a}o
baseada na mec\^{a}nica qu\^{a}ntica n\~{a}o-relativ\'{\i}stica, tanto para
uma barreira de potencial quanto para um po\c{c}o de potencial, e que s\'{o}
podem ser explicados pela media\c{c}\~{a}o de antipart\'{\i}culas. A an\'{a}%
lise do efeito t\'{u}nel relativ\'{\i}stico nos conduz \`{a} conclus\~{a}o
que o paradoxo de Klein, usualmente relacionado com a emiss\~{a}o de antipart%
\'{\i}culas por potenciais suficientemente intensos, n\~{a}o \'{e}
necessariamente relacionado com a irradia\c{c}\~{a}o de antipart\'{\i}culas.
Isto dito tendo em vista que, apesar da forma\c{c}\~{a}o de estados ligados
de antipart\'{\i}culas na regi\~{a}o de intera\c{c}\~{a}o, induzida pelas
part\'{\i}culas incidentes, somente part\'{\i}culas s\~{a}o, de fato,
irradiadas. A investiga\c{c}\~{a}o dos estados ligados permite-nos concluir
sobre a influ\^{e}ncia de um contaminante escalar no efeito SSW. Verifica-se
que a presen\c{c}a de um acoplamento escalar inibe o efeito SSW, e o torna
invi\'{a}vel se o acoplamento escalar exceder o acoplamento vetorial.

Apesar da originalidade e generalidade, este trabalho \'{e} acess\'{\i}vel
aos estudantes de gradua\c{c}\~{a}o em f\'{\i}sica que tenham freq\"{u}%
entado alguns poucos meses de um curso introdut\'{o}rio de mec\^{a}nica qu%
\^{a}ntica. Dessa forma permite-se o acesso precoce de estudantes a alguns
dos mais interessantes fen\^{o}menos da MQR.

\section{A equa\c{c}\~{a}o de Klein-Gordon}

A EKG unidimensional para uma part\'{\i}cula livre de massa de repouso $m$
cor\-res\-pon\-de \`{a} rela\c{c}\~{a}o energia-momento relativ\'{\i}stica $%
E^{2}=c^{2}p^{2}+m^{2}c^{4}$, onde a energia $E$ e o momento $p$ tornam-se
o\-pe\-ra\-do\-res, $i\hbar \,\partial /\partial t$ e $-i\hbar \,\partial
/\partial x $ respectivamente, atuando sobre a fun\c{c}\~{a}o de onda $\Phi
(x,t)$. Aqui, $c$ \'{e} a velocidade da luz e $\hbar $ \'{e} a constante de
Planck ($\hbar =h/(2\pi )$).

Na presen\c{c}a de potenciais externos a rela\c{c}\~{a}o energia-momento
torna-se%
\begin{equation}
\left( E-V_{t}\right) ^{2}=c^{2}\left( p-\frac{V_{e}}{c}\right) ^{2}+\left(
mc^{2}+V_{s}\right) ^{2}  \label{1}
\end{equation}%
onde os subscritos nos termos dos potenciais denotam suas propriedades com
res\-pei\-to \`{a}s transforma\c{c}\~{o}es de Lorentz: $t$ e $e$ para os
componentes temporal e espacial de um potencial vetorial\footnote{%
A energia e o momento s\~{a}o os componentes temporal e espacial,
respectivamente, da quantidade $(E/c\,,\,p)$, a qual se comporta, segundo as
transforma\c{c}\~{o}es de Lorentz, como um vetor. O potencial vetorial, com
componentes $(V_{t}\,,\,V_{e})$, \'{e} acoplado \`{a} part\'{\i}cula de
acordo com o \textit{princ\'{\i}pio do acoplamento m\'{\i}nimo}, tamb\'{e}m
chamado de \textit{princ\'{\i}pio da substitui\c{c}\~{a}o m\'{\i}nima}, $%
E\rightarrow E-V_{t}$ e $p\rightarrow p-V_{e}/c$, como \'{e} habitual no
caso da intera\c{c}\~{a}o eletromagn\'{e}tica.}, e $s$ para um potencial
escalar\footnote{%
A massa de repouso \'{e} uma quantidade invariante de Lorentz, i. e., uma
quantidade escalar. O potencial escalar foi acoplado \`{a} part\'{\i}cula em
(1) de acordo com o \textit{princ\'{\i}pio do acoplamento m\'{\i}nimo} $%
m\rightarrow m+V_{s}/c^{2}$. Esta prescri\c{c}\~{a}o fornece o limite n\~{a}%
o-relativ\'{\i}stico apropriado da EKG, conforme Ref. \cite{tati}, em
contraste com a regra $m^{2}\rightarrow m^{2}+V_{s}^{2}/c^{4}$ empregada na
Ref. \cite{gre}.}.

A equa\c{c}\~{a}o da continuidade para a EKG%
\begin{equation}
\frac{\partial \rho }{\partial t}+\frac{\partial J}{\partial x}=0
\label{con}
\end{equation}%
\noindent \'{e} satisfeita com $\rho $ e $J$ definidos como%
\begin{equation}
\rho =\frac{i\hbar }{2mc^{2}}\left( \Phi ^{\ast }\frac{\partial \Phi }{%
\partial t}-\frac{\partial \Phi ^{\ast }}{\partial t}\Phi \right) -\frac{%
V_{t}}{mc^{2}}\left\vert \Phi \right\vert ^{2}  \label{rho}
\end{equation}%
\begin{equation}
J=\frac{\hbar }{2im}\left( \Phi ^{\ast }\frac{\partial \Phi }{\partial x}-%
\frac{\partial \Phi ^{\ast }}{\partial x}\Phi \right) -\frac{V_{e}}{mc}%
\left\vert \Phi \right\vert ^{2}  \label{jota}
\end{equation}%
Vale a pena observar o modo que os componentes do potencial vetorial
participam da densidade $\rho $ e da corrente $J$, tanto quanto a aus\^{e}%
ncia do potencial escalar. Observa-se tamb\'{e}m que a densidade envolve
derivadas temporais, um fato relacionado com a derivada temporal de segunda
ordem na EKG, e pode admitir valores negativos mesmo no caso de uma part%
\'{\i}cula livre. Assim sendo $\rho $ n\~{a}o pode ser interpretada como uma
densidade de probabilidade. Contudo, Pauli e Weisskopf \cite{pau} mostraram
que n\~{a}o h\'{a} dificuldade com a interpreta\c{c}\~{a}o da densidade e da
corrente da EKG se essas grandezas forem interpretadas como densidade e
corrente de \textit{carga}, ao inv\'{e}s de densidade e corrente de
probabilidade. A \textit{carga}\ n\~{a}o deve ser pensada necessariamente
como carga el\'{e}trica, mas como carga ge\-ne\-ra\-li\-za\-da que satisfaz
uma lei de conserva\c{c}\~{a}o aditiva, por assim dizer que a \textit{carga}
de um sistema \'{e} a soma das \textit{cargas} de suas partes constituintes.

Para potenciais externos independentes do tempo, a EKG admite solu\c{c}\~{o}%
es da forma%
\begin{equation}
\Phi (x,t)=\phi (x)\,e^{i\Lambda \left( x\right) }\,e^{-i\frac{E}{\hbar }t}
\label{2a}
\end{equation}

\noindent onde $\phi $ obedece a uma equa\c{c}\~{a}o similar em forma \`{a}
equa\c{c}\~{a}o de Schr\"{o}dinger

\begin{equation}
-\frac{\hbar ^{2}}{2m}\,\frac{d^{2}\phi }{dx^{2}}+\left( \frac{%
V_{s}^{2}-V_{t}^{2}}{2mc^{2}}+V_{s}+\frac{E}{mc^{2}}\,V_{t}\right) \,\phi =%
\frac{E^{2}-m^{2}c^{4}}{2mc^{2}}\,\phi  \label{3}
\end{equation}%
com $\Lambda \left( x\right) =\int^{x}dy\,V_{e}(y)/(\hbar c)$. A elimina\c{c}%
\~{a}o do componente espacial do potencial vetorial \'{e} e\-qui\-va\-len%
\-te a uma redefini\c{c}\~{a}o do operador momento. Realmente,
\begin{equation}
\left( p_{op}-\frac{V_{e}}{c}\right) ^{2}\Phi =e^{i\Lambda
}\,p_{op}^{2}\,\phi  \label{mom}
\end{equation}

\'{E} agora importante perceber que h\'{a} solu\c{c}\~{o}es de energia
positiva tanto quanto solu\c{c}\~{o}es de energia negativa\footnote{%
As solu\c{c}\~{o}es de energia positiva e negativa s\~{a}o associadas com
part\'{\i}culas e antipart\'{\i}culas, respectivamente.} e que os dois poss%
\'{\i}veis sinais para $E$ implicam em duas possibilidades para a evolu\c{c}%
\~{a}o temporal da fun\c{c}\~{a}o de onda. Seja como for, a energia \'{e}
uma quantidade conservada. A forma da equa\c{c}\~{a}o de autovalor (\ref{3})
\'{e} preservada sob as transforma\c{c}\~{o}es simult\^{a}neas $E\rightarrow
-E$ e $V_{t}\rightarrow -V_{t}$, e isto implica que part\'{\i}culas e
antipart\'{\i}culas est\~{a}o sujeitas a componentes temporais de um
potencial vetorial com sinais dissimilares. Como conseq\"{u}\^{e}ncia
imediata dessa covari\^{a}ncia tem-se que, \ por mais estranho que possa
parecer, part\'{\i}culas e antipart\'{\i}culas compartilham exatamente a
mesma autofun\c{c}\~{a}o no caso de um potencial puramente escalar e que o
espectro \'{e} disposto simetricamente em torno de $E=0$. \textit{Cargas}
positivas e negativas est\~{a}o sujeitas a acoplamentos vetoriais
(componentes temporais) de sinais contr\'{a}rios e igual acoplamento
escalar. A intera\c{c}\~{a}o escalar \'{e} independente da \textit{carga} e
assim age indiscriminadamente sobre part\'{\i}culas e antipart\'{\i}culas.
Diz-se ent\~{a}o que o potencial vetorial acopla com a \textit{carga} da part%
\'{\i}cula e que o potencial escalar acopla com a massa da part\'{\i}cula. \
A densidade e a corrente correspondentes \`{a} solu\c{c}\~{a}o expressa por (%
\ref{2a}) tornam-se%
\begin{equation}
\rho =\frac{E-V_{t}}{mc^{2}}\left\vert \phi \right\vert ^{2}  \label{rho1}
\end{equation}%
\begin{equation}
J=\frac{\hbar }{2im}\left( \phi ^{\ast }\frac{\partial \phi }{\partial x}-%
\frac{\partial \phi ^{\ast }}{\partial x}\phi \right)  \label{jota1}
\end{equation}%
Em virtude de $\rho $ e $J$ serem independentes do tempo, a solu\c{c}\~{a}o (%
\ref{2a}) \'{e} dita descrever um estado estacion\'{a}rio. Nota-se que a
densidade torna-se negativa em regi\~{o}es do espa\c{c}o onde $V_{t}>E$ e
que o componente espacial do potencial vetorial n\~{a}o mais interv\'{e}m na
corrente. Tamb\'{e}m, a lei de conserva\c{c}\~{a}o da carga dada por (\ref%
{con}) implica que a corrente \'{e} independente de $x$ para os estados
estacion\'{a}rios.

\section{Solu\c{c}\~{a}o para um potencial quadrado}

Vamos agora considerar a EKG com os potenciais externos independentes do
tempo na forma de um potencial quadrado. Consideramos $V_{e}=0$, haja vista
que o componente espacial do potencial vetorial contribui apenas com um
fator de fase local para $\Phi (x,t)$ e n\~{a}o contribui para a densidade
nem para a corrente. O potencial quadrado \'{e} expresso como
\begin{equation}
V(x)=V_{0}\left[ \theta \left( x+a\right) -\theta \left( x-a\right) \right]
=\left\{
\begin{array}{c}
0 \\
\\
V_{0}%
\end{array}%
\begin{array}{c}
{\textrm{para }}|x|>a \\
\\
{\textrm{para }}|x|<a%
\end{array}%
\right.  \label{4}
\end{equation}%
onde $a>0$ e $\theta (x)$ \'{e} a fun\c{c}\~{a}o de Heaviside. $V_{0}>0$
para uma barreira de potencial e $V_{0}<0$ para um po\c{c}o de potencial. Os
potenciais vetorial e escalar s\~{a}o escritos como $V_{t}(x)=g_{t}V(x)$ e $%
V_{s}(x)=g_{s}V(x)$ de tal forma que as constantes de acoplamento est\~{a}o
sujeitas ao v\'{\i}nculo $g_{t}+g_{s}=1$, com $g_{t}\geq 0$ e $g_{s}\geq 0$.

Para $x<-a$, a EKG apresenta a solu\c{c}\~{a}o geral%
\begin{equation}
\phi =A_{+}\,e^{+ikx}+A_{-}\,e^{-ikx}  \label{5}
\end{equation}%
onde
\begin{equation}
k=\frac{\sqrt{E^{2}-m^{2}c^{4}}}{\hbar c}  \label{6}
\end{equation}%
Para $|E|>mc^{2}$, a solu\c{c}\~{a}o expressa por (\ref{5}) reverte-se uma
soma de autofun\c{c}\~{o}es do operador momento e descrevem ondas planas
pro\-pa\-gan\-do-se em ambos os sentidos do eixo $X$ com velocidade de grupo%
\footnote{%
Veja, e.g., Refs. [1] e [5].}%
\begin{equation}
v_{g}=\frac{1}{\hbar }\,\frac{dE}{dk}  \label{7}
\end{equation}%
igual \`{a} velocidade cl\'{a}ssica da part\'{\i}cula. Se escolhermos part%
\'{\i}culas incidindo sobre a regi\~{a}o de potencial ($E>mc^{2}$) teremos
que $A_{+}\,e^{+ikx}$ descreve par\-t\'{\i}\-cu\-las incidentes ($%
v_{g}=c^{2}\hbar k/E>0$), enquanto $A_{-}\,e^{-ikx}$ descreve part\'{\i}%
culas re\-fle\-ti\-das ($v_{g}=-c^{2}\hbar k/E<0$). $\ $ Os pap\'{e}is das
ondas ser\~{a}o invertidos se con\-si\-de\-rar\-mos a incid\^{e}ncia de
antipart\'{\i}culas. Doravante, por motivos de simplicidade e sem perda de
generalidade, consideraremos apenas a incid\^{e}ncia de part\'{\i}culas. A
corrente nesta regi\~{a}o do espa\c{c}o, correspondendo a $\phi $ dada por \
(\ref{5}), \'{e} expressa por
\begin{equation}
J=J_{\mathtt{inc}}-J_{\mathtt{ref}}  \label{7a}
\end{equation}%
onde%
\begin{equation}
J_{\mathtt{inc}}=\frac{\hbar k}{m}|A_{+}|^{2},\quad J_{\mathtt{ref}}=\frac{%
\hbar k}{m}|A_{-}|^{2}  \label{70a}
\end{equation}%
Observe que a rela\c{c}\~{a}o $J=\rho \,v_{g}$ mant\'{e}m-se tanto para a
onda incidente quanto para a onda refletida pois%
\begin{equation}
\rho _{\pm }=\frac{E}{mc^{2}}\,|A_{\pm }|^{2}  \label{7b}
\end{equation}

Por outro lado, para $x>a$ as solu\c{c}\~{o}es s\~{a}o da forma%
\begin{equation}
\phi =C_{+}\,e^{+ikx}+C_{-}\,e^{-ikx}  \label{7c}
\end{equation}%
Para termos uma onda progressiva se afastando da regi\~{a}o do potencial
(pro\-pa\-gan\-do-se no sentido positivo do eixo $X$ com $v_{g}=c^{2}\hbar
k/E>0$) devemos impor $C_{-}=0$. A densidade e a corrente nesta regi\~{a}o
do espa\c{c}o, correspondendo a $\phi $ dada por \ (\ref{7c}) com $C_{-}=0$,
s\~{a}o expressas por%
\begin{equation}
\rho =\frac{E}{mc^{2}}\,|C_{+}|^{2},\quad J_{\mathtt{trans}}=\frac{\hbar k}{m%
}|C_{+}|^{2}  \label{70c}
\end{equation}

Note que essas solu\c{c}\~{o}es para $|x|>a$ descrevem estados de
es\-pa\-lha\-men\-to com $|E|>mc^{2}$ e $k$ $\in
\mathbb{R}
$. Poss\'{\i}veis estados ligados tamb\'{e}m poderiam ser descritos por (\ref%
{5}) e (\ref{7c}) com $k=i\kappa $, onde $\kappa =\allowbreak \sqrt{%
m^{2}c^{4}-E^{2}}/(\hbar c)$ com $|E|<mc^{2}$, e $A_{+}=C_{-}=0$.

Para $-a<x<a$ a solu\c{c}\~{a}o geral tem a forma%
\begin{equation}
\phi =B_{+}\,e^{+iqx}+B_{-}\,e^{-iqx}  \label{8}
\end{equation}%
onde
\begin{equation}
q=\frac{\sqrt{\left( E-g_{t}V_{0}\right) ^{2}-\left(
mc^{2}+g_{s}V_{0}\right) ^{2}}}{\hbar c}  \label{9}
\end{equation}%
As solu\c{c}\~{o}es $B_{\pm }\,e^{\pm iqx}$ com $q\in
\mathbb{R}
$ descrevem ondas planas que se propagam com velocidade de grupo%
\begin{equation}
v_{g}=\pm \frac{c^{2}\hbar q}{E-g_{t}V_{0}}  \label{9a}
\end{equation}%
com densidades e correntes associadas dadas por
\begin{equation}
\rho _{\pm }=\frac{E-g_{t}V_{0}}{mc^{2}}\,|B_{\pm }|^{2},\quad J_{\pm }=%
\frac{\hbar q}{m}|B_{\pm }|^{2}  \label{11a}
\end{equation}%
Na circunst\^{a}ncia em que $E<g_{t}V_{0}$ nos defrontamos com um caso
bizarro, pois as densidades s\~{a}o quantidades negativas. A manten\c{c}a da
rela\c{c}\~{a}o $J=\rho \,v_{g}$ para cada solu\c{c}\~{a}o particular,
contudo, \'{e} uma licen\c{c}a para interpretar $B_{+}\,e^{+iqx}$ ($%
B_{-}\,e^{-iqx}$) a descrever a propaga\c{c}\~{a}o, no sentido negativo
(positivo) do eixo $X$, de part\'{\i}culas com \textit{carga}\ de sinal contr%
\'{a}rio ao das part\'{\i}culas incidentes\footnote{%
Isto torna-se poss\'{\i}vel por causa da dupla possibilidade de sinais para
a energia de um estado estacion\'{a}rio. Acontece que $B_{\pm }\,e^{\pm iqx}$
\ pode vir a descrever uma onda progressiva \ com energia negativa e
velocidade de fase $v_{f}=\mp |E|/(\hbar c)$. A car\^{e}ncia da interpreta%
\c{c}\~{a}o f\'{\i}sica deste resultado ser\'{a} suprida numa se\c{c}\~{a}o
posterior.}. A corrente na regi\~{a}o $|x|<a$, para \textit{cargas}
positivas tanto como para \textit{cargas} negativas, pode ent\~{a}o ser
escrita como%
\begin{equation}
J=\frac{\hbar q}{m}\left( |B_{+}|^{2}-|B_{-}|^{2}\right)  \label{11b}
\end{equation}%
No caso em que $q=i|q|$ n\~{a}o h\'{a} ondas progressivas na regi\~{a}o do
potencial. Contudo h\'{a} ainda uma corrente dada por%
\begin{equation}
J=\frac{i\hbar |q|}{m}\left( B_{+}B_{-}^{\ast }-B_{+}^{\ast }B_{-}\right)
\label{9e}
\end{equation}

\section{Espalhamento}

Vamos considerar a incid\^{e}ncia de part\'{\i}culas ($E>mc^{2}$) e assim $k$%
, definido em (\ref{6}), \'{e} uma quantidade real. A discrimina\c{c}\~{a}o
entre $q$ real e $q$ imagin\'{a}rio permite-nos identificar dois valores cr%
\'{\i}ticos para o potencial%
\begin{equation}
V_{1}=E-mc^{2},\quad \textrm{e\quad }V_{2}=\frac{E+mc^{2}}{2g_{t}-1}\quad
\textrm{com\quad }g_{t}\neq 1/2  \label{9f}
\end{equation}%
e o valor de $g_{t}$ nos permite segregar tr\^{e}s classes distintas de solu%
\c{c}\~{o}es de es\-pa\-lha\-men\-to:

\medskip

\begin{itemize}
\item \textbf{Classe A - }$\mathbf{g}_{t}\,\mathbf{>1/2}$\textbf{. } Nesta
classe temos que $q\in
\mathbb{R}
$ para $V_{0}<V_{1}$ e $V_{0}>V_{2}$ (note que $V_{2}>0$). Ondas n\~{a}%
o-progressivas, correspondendo a $q=i|q|$, ocorrem para uma barreira de
potencial de altura mediana ($V_{1}<V_{0}<V_{2}$). Um c\'{a}lculo simples
revela que $E>g_{t}V_{0}$ para $V_{0}<V_{1}$. Entretanto, $E<g_{t}V_{0}$
para $V_{0}>V_{2}$, a circunst\^{a}ncia em que h\'{a} a propaga\c{c}\~{a}o
de \textit{cargas} negativas confinadas na regi\~{a}o $|x|<a$.

\item \textbf{Classe B - }$\mathbf{g}_{t}\,\mathbf{=1/2}$\textbf{. } Apenas $%
V_{1}$ desempenha papel na distin\c{c}\~{a}o entre valores reais e imagin%
\'{a}rios do n\'{u}mero de onda $q$. Aqui $q\in
\mathbb{R}
$ somente para $V_{0}<V_{1}$. H\'{a} apenas solu\c{c}\~{o}es com $%
E>g_{t}V_{0}$ e ondas n\~{a}o-progressivas ocorrem para uma barreira de
potencial tanto ou quanto alta ($V_{0}>V_{1}$).

\item \textbf{Classe C - }$\mathbf{g}_{t}\,\mathbf{<1/2}$\textbf{. } Temos
que $q\in
\mathbb{R}
$ para $V_{2}<V_{0}<V_{1}$ (note que $V_{2}<0$). N\~{a}o mais que solu\c{c}%
\~{o}es com $E>g_{t}V_{0}$ s\~{a}o permitidas e ondas evanescentes ocorrem
para uma barreira de potencial suficientemente alta ($V_{0}>V_{1}$). O que
causa estranheza \'{e} que ondas evanescentes v\^{e}m \`{a} tona para um po%
\c{c}o de potencial um tanto profundo ($V_{0}<V_{2}$).
\end{itemize}

Come\c{c}aremos agora o c\'{a}lculo de grandezas de suma import\^{a}ncia na
des\-cri\-\c{c}\~{a}o do espalhamento, viz., os coeficientes de reflex\~{a}o
e transmiss\~{a}o. N\~{a}o obstante a descontinuidade do potencial em $x=\pm
a$, a autofun\c{c}\~{a}o e sua derivada primeira s\~{a}o fun\c{c}\~{o}es cont%
\'{\i}nuas\footnote{%
Esta conclus\~{a}o, v\'{a}lida para potenciais com descontinuidades finitas,
pode ser obtida pela integra\c{c}\~{a}o da Eq. (\ref{3}) entre $-\varepsilon
$ e $+\varepsilon $ no limite $\varepsilon \rightarrow 0$. Pode-se
verificar, pelo mesmo pro\-ce\-di\-men\-to, que apenas as autofun\c{c}\~{o}%
es s\~{a}o cont\'{\i}nuas quando as descontinuidades dos potenciais s\~{a}o
infinitas.}. A demanda por continuidade de $\phi $ e $d\phi /dx$ fixa todas
as amplitudes em termos da amplitude da onda incidente $A_{+}$, viz.

\begin{equation}
\frac{A_{-}}{A_{+}}=\frac{i(q^{2}-k^{2})\sin (2qa)\exp (-2ika)}{2kq\cos
(2qa)-i(q^{2}+k^{2})\sin (2qa)}  \label{13a}
\end{equation}

\begin{equation}
\frac{B_{+}}{A_{+}}=\frac{k\left( q+k\right) \exp \left[ -i\left( q+k\right)
a\right] }{2kq\cos (2qa)-i(q^{2}+k^{2})\sin (2qa)}  \label{13b}
\end{equation}

\begin{equation}
\frac{B_{-}}{A_{+}}=\frac{k\left( q-k\right) \exp \left[ +i\left( q-k\right)
a\right] }{2kq\cos (2qa)-i(q^{2}+k^{2})\sin (2qa)}  \label{13c}
\end{equation}

\begin{equation}
\frac{C_{+}}{A_{+}}=\frac{2qk\exp (-2ika)}{2qk\cos (2qa)-i(q^{2}+k^{2})\sin
(2qa)}  \label{13d}
\end{equation}%
\noindent onde as identidades matem\'{a}ticas $\sin \left( i\theta \right)
=i\sinh \left( \theta \right) $ e $\cos \left( i\theta \right) =$ cosh$%
\left( \theta \right) $ foram utilizadas para escrever as amplitudes
relativas para os casos de $q$ real e imagin\'{a}rio puro numa forma compacta%
\footnote{%
A transi\c{c}\~{a}o de $q$ real para $q$ imagin\'{a}rio puro \'{e} feita
pela prescri\c{c}\~{a}o $q\rightarrow i|q|$.}.

Agora focalizamos nossa aten\c{c}\~{a}o na determina\c{c}\~{a}o dos
coeficientes de reflex\~{a}o $R$ e transmiss\~{a}o $T$. O coeficiente de
reflex\~{a}o (transmiss\~{a}o) \'{e} definido como a raz\~{a}o entre as
correntes refletida (transmitida) e incidente. Haja vista que $\partial \rho
/\partial t=0$ para estados estacion\'{a}rios, temos que a corrente \'{e}
independente de $x$. Usando este fato obtemos prontamente que

\begin{equation}
R=\frac{|A_{-}|^{2}}{|A_{+}|^{2}}=\left\{ 1+\left[ \frac{2qk}{\left(
k^{2}-q^{2}\right) \sin (2qa)}\right] ^{2}\right\} ^{-1}  \label{15}
\end{equation}

\begin{equation}
T=\frac{|C_{+}|^{2}}{|A_{+}|^{2}}=\left\{ 1+\left[ \frac{k^{2}-q^{2}}{2qk}%
\sin (2qa)\right] ^{2}\right\} ^{-1}  \label{16}
\end{equation}%
Seja l\'{a} como for temos que $R+T=1$, como deve ser por causa da conserva%
\c{c}\~{a}o da \textit{carga}.

Ademais, deve ser observado que quando as ondas progressivas para a direita
e para a esquerda \ na regi\~{a}o do potencial ($q\in
\mathbb{R}
$) interferem de modo a formar uma onda estacion\'{a}ria (i.e., quando $%
2qa=n\pi $ com $n=1,2,3,\ldots $), sucede a transmiss\~{a}o ressonante ($T=1$%
), correspondendo \`{a}s energias dadas por%
\begin{equation}
E_{n}=g_{t}V_{0}\pm \sqrt{\left( \frac{n\pi \hbar c}{2a}\right) ^{2}+\left(
mc^{2}+g_{s}V_{0}\right) ^{2}}  \label{160}
\end{equation}%
onde o sinal defronte do radical deve ser escolhido de forma que $E>mc^{2}$.
Note que a resson\^{a}ncia de transmiss\~{a}o pode ocorrer tanto para uma
barreira de potencial quanto para um po\c{c}o de potencial.

As Figuras 1, 2 e 3 ilustram o coeficiente de transmiss\~{a}o para as tr\^{e}%
s classes de solu\c{c}\~{o}es dis\-cri\-mi\-na\-das anteriormente. Em todas
as figuras usamos $E=1,1\,mc^{2}$ e consideramos $a/\lambda $ \'{e} igual a
1 nas Figuras 1 e 2, e $a/\lambda =3$ na Figura 3. Usamos o sistema de
unidades em que $\hbar =c=m=1$.

A Classe A apresenta um comportamento an\^{o}malo para uma barreira de
potencial muito alta. Vimos anteriormente que para $g_{t}>1/2$ e $%
V_{0}>V_{2} $ h\'{a} ondas progressivas na regi\~{a}o do potencial enquanto
a mec\^{a}nica qu\^{a}ntica n\~{a}o-relativ\'{\i}stica prev\^{e} a e\-xis\-t%
\^{e}n\-cia de ondas evanescentes. De mais a mais, a mec\^{a}nica qu\^{a}%
ntica n\~{a}o-relativ\'{\i}stica profetiza que, neste caso de tunelamento, o
coeficiente de transmiss\~{a}o sofre uma supress\~{a}o exponencial \`{a}
medida que o potencial tende ao infinito, enquanto que nosso resultado
prenuncia um coeficiente de transmiss\~{a}o oscilat\'{o}rio que e\-xi\-be at%
\'{e} mesmo um comportamento ressonante. A Classe B n\~{a}o apresenta
resultados que contradizem a intui\c{c}\~{a}o baseada na mec\^{a}nica qu\^{a}%
ntica n\~{a}o-relativ\'{\i}stica. A Classe C, entretanto, e\-xi\-be um
comportamento an\^{o}malo para um po\c{c}o de potencial com $V_{0}<V_{2}$: a
supress\~{a}o exponencial do coeficiente de transmiss\~{a}o \`{a} medida que
o po\c{c}o de potencial torna-se muito profundo.

Da discuss\~{a}o relacionada com a Classe A, observa-se que o limiar para a
e\-xis\-t\^{e}n\-cia de \textit{cargas} negativas na regi\~{a}o do potencial
\'{e} dado por $V_{2}$. Pode-se interpretar a possibilidade de propaga\c{c}%
\~{a}o de \textit{cargas} negativas na regi\~{a}o da barreira de potencial
como sendo de\-vido ao fato que \ cada \textit{carga} negativa tem energia $%
-E$ e est\'{a} sujeita a um potencial efetivo dado por $\left(
g_{s}-g_{t}\right) V_{0}$. Quer dizer, ent\~{a}o, que a ondas progressivas
nessa regi\~{a}o de potencial des\-cre\-vem, de fato, a propaga\c{c}\~{a}o
de antipart\'{\i}culas\footnote{%
Note que part\'{\i}cula e antipart\'{\i}cula t\^{e}m massas iguais.}.
Destarte se $g_{t}>1/2$ a antipart\'{\i}cula ter\'{a} uma energia dispon%
\'{\i}vel (energia de repouso mais energia cin\'{e}tica) expressa por $%
-E-\left( 1-2g_{t}\right) V_{0}$, donde se conclui sobre a energia do limiar
para a exist\^{e}ncia de \textit{cargas} ne\-ga\-ti\-vas (antipart\'{\i}%
culas) movendo-se na regi\~{a}o do potencial. Nas regi\~{o}es com $|x|>a$
com $V_{0}>V_{2}$, tais antipart\'{\i}culas ser\~{a}o descritas por ondas
evanescentes pois o potencial efetivo torna-se maior que $-E$. \ As part%
\'{\i}culas est\~{a}o sob a influ\^{e}ncia de um potencial dado $\left(
g_{s}+g_{t}\right) V_{0}=V_{0}$, e se as part\'{\i}culas estiverem sujeitas
a uma barreira de potencial ent\~{a}o as antipart\'{\i}culas estar\~{a}o
sujeitas a um \ po\c{c}o (barreira) de potencial se $g_{t}>1/2$ ($g_{t}<1/2$%
). \'{E} razo\'{a}vel conjecturar que as part\'{\i}culas incidentes induzem
a forma\c{c}\~{a}o de estados ligados de antipart\'{\i}culas na regi\~{a}o
de intera\c{c}\~{a}o. Esta id\'{e}ia \'{e} refor\c{c}ada pela observa\c{c}%
\~{a}o que, no caso de $V_{0}>V_{2}$ e $g_{t}>1/2$, as antipart\'{\i}culas
est\~{a}o sujeitas a um po\c{c}o de potencial efetivo com energia que excede
o m\'{\i}nimo do potencial (fundo do po\c{c}o) e \'{e} menor que o potencial
m\'{a}ximo (boca do po\c{c}o). Mais ainda, para $V_{0}<V_{2}$ e $g_{t}>1/2$,
a e\-ner\-gi\-a da antipart\'{\i}cula \'{e} menor que o valor m\'{\i}nimo do
potencial efetivo, uma circunst\^{a}ncia bem conhecida em que a autofun\c{c}%
\~{a}o n\~{a}o satisfaz as condi\c{c}\~{o}es de contorno apropriadas, ora
pois tal solu\c{c}\~{a}o n\~{a}o \'{e} aceit\'{a}vel.

A situa\c{c}\~{a}o paradoxal observada no espalhamento da Classe C, para um
po\c{c}o suficientemente profundo ($V_{0}<V_{2}<0$), tamb\'{e}m pode ser
elucidada com a argumenta\c{c}\~{a}o delineada no par\'{a}grafo anterior.
Realmente, quando $g_{t}<1/2$ e a part\'{\i}cula est\'{a} sujeita a um po%
\c{c}o de potencial, h\'{a} um po\c{c}o de potencial efetivo para a antipart%
\'{\i}cula com uma energia dispon\'{\i}vel dada por $-E-\left(
2g_{t}-1\right) |V_{0}|$. L\'{a} ent\~{a}o torna-se poss\'{\i}vel a propaga%
\c{c}\~{a}o de antipart\'{\i}culas. H\'{a} a propaga\c{c}\~{a}o prom\'{\i}%
scua de \textit{cargas} positivas (part\'{\i}culas) e negativas\textit{\ \ }%
(antipart\'{\i}culas) na regi\~{a}o do potencial, com uma densidade de
\textit{carga} resultante positiva quando o po\c{c}o de potencial torna-se
muito profundo e $g_{t}<1/2$. Na regi\~{a}o externa do potencial ($|x|>a$)
tais antipart\'{\i}culas ser\~{a}o descritas por ondas evanescentes j\'{a}
que o potencial efetivo \'{e} maior que $-E$. Aqui, tal como na Classe A, h%
\'{a} forma\c{c}\~{a}o de estados ligados de antipart\'{\i}culas induzida
pela incid\^{e}ncia de part\'{\i}culas.

Em todas as circunst\^{a}ncias, consideramos a incid\^{e}ncia de part\'{\i}%
culas e obtivemos como resultado a transmiss\~{a}o de part\'{\i}culas. Em
determinadas situa\c{c}\~{o}es de potenciais extremamentes intensos,
entretanto, chegamos ao entendimento que h\'{a} estados ligados de antipart%
\'{\i}culas na regi\~{a}o do potencial. H\'{a} uma pergunta que n\~{a}o pode
calar: qual a origem de tais antipart\'{\i}culas?

Para $g_{t}>1/2$ e $V_{0}>V_{2}>0$, os \textit{continua} com $E>mc^{2}$ para
as part\'{\i}culas e $E<-mc^{2}$ para as antipart\'{\i}culas que existe para
$|x|>a$ tornam-se $E>mc^{2}+V_{0}$ para as part\'{\i}culas e $%
E<-mc^{2}-\left( g_{s}-g_{t}\right) V_{0}$ para as antipart\'{\i}culas em $%
|x|<a$. Quando uma part\'{\i}cula incide em $x=-a$ com energia menor que $%
-mc^{2}-\left( g_{s}-g_{t}\right) V_{0}$, o componente vetorial da barreira
de potencial estimula a produ\c{c}\~{a}o de antipart\'{\i}culas. Em virtude
da conserva\c{c}\~{a}o da \textit{carga}\ h\'{a}, em verdade, a cria\c{c}%
\~{a}o de pares part\'{\i}cula-antipart\'{\i}cula e, como o potencial
vetorial em $|x|<a$ \'{e} repulsivo para part\'{\i}culas e atrativo para
antipart\'{\i}culas, as part\'{\i}culas ser\~{a}o necessariamente ejetadas
da regi\~{a}o (para a esquerda) e as antipart\'{\i}culas ser\~{a}o
necessariamente confinadas na regi\~{a}o do potencial. Quando uma antipart%
\'{\i}cula criada em $x=-a$ incide em $x=+a$ o componente vetorial da
barreira de potencial estimula mais uma vez a produ\c{c}\~{a}o de pares,
agora ent\~{a}o, as part\'{\i}culas ser\~{a}o ejetadas para a direita da regi%
\~{a}o de intera\c{c}\~{a}o. N\~{a}o apenas a\ \textit{carga}\ \'{e}
conservada no processo de cria\c{c}\~{a}o de pares. Visto que os pares
produzidos em $x=\pm a$ t\^{e}m energias de sinais contr\'{a}rios,
conclui-se que a energia tamb\'{e}m \'{e} uma quantidade conservada no
processo de cria\c{c}\~{a}o de pares. Apesar da possibilidade da cria\c{c}%
\~{a}o de pares part\'{\i}cula-antipart\'{\i}cula nas interfaces do
potencial localizadas em $x=\pm a$, um fen\^{o}meno relacionado com o
paradoxo de Klein, \ e a propaga\c{c}\~{a}o de antipart\'{\i}culas na regi%
\~{a}o $|x|<a,$ a conjuntura n\~{a}o \'{e} prop\'{\i}cia \`{a} emiss\~{a}o
de antipart\'{\i}culas. Poder\'{\i}amos ent\~{a}o nos aventurar a afirmar
que o paradoxo de Klein \'{e} inerente \`{a}s equa\c{c}\~{o}es relativ\'{\i}%
sticas e n\~{a}o \'{e} necessariamente relacionado com a irradia\c{c}\~{a}o
de antipart\'{\i}culas, como \'{e} propalado na literatura. Torna-se
evidente que o acoplamento escalar resulta no aumento da energia m\'{\i}nima
necess\'{a}ria para a cria\c{c}\~{a}o de pares part\'{\i}cula-antipart\'{\i}%
cula. O valor m\'{\i}nimo do limiar ($V_{0}=2mc^{2})$ ocorre quando o
acoplamento \'{e} puramente vetorial ($g_{t}=1$). A adi\c{c}\~{a}o de um
contaminante escalar contribui para aumentar o valor do limiar, o qual,
surpreendentemente, torna-se infinito j\'{a} para uma mistura meio a meio de
acoplamentos. Deste modo, a produ\c{c}\~{a}o de pares n\~{a}o \'{e} fact%
\'{\i}vel se o acoplamento vetorial n\~{a}o exceder o acoplamento escalar,
ainda que a barreira de potencial $V_{0}$ seja extremamente alta.

Para $g_{t}<1/2$ e $V_{0}<V_{2}<0$, por\'{e}m, os \textit{continua} com $%
E>mc^{2}$ para as part\'{\i}culas e $E<-mc^{2}$ para as antipart\'{\i}culas
que existe para $|x|>a$ tornam-se $E>mc^{2}-|V_{0}|$ para as part\'{\i}culas
e $E<-mc^{2}+\left( g_{s}-g_{t}\right) |V_{0}|$ para as antipart\'{\i}culas
em $|x|<a$. Neste caso, quando a part\'{\i}cula incidente tem energia menor
que $-mc^{2}+\left( g_{s}-g_{t}\right) |V_{0}|$ part\'{\i}culas e antipart%
\'{\i}culas propagam-se efetivamente num po\c{c}o de potencial. Conv\'{e}m
lembrar que as antipart\'{\i}culas n\~{a}o t\^{e}m chance de se propagar
para as regi\~{o}es externas do po\c{c}o de potencial, e assim sendo, tal
como no caso com $g_{t}>1/2$, est\~{a}o confinadas \`{a} regi\~{a}o $|x|<a$.
\'{E} merit\'{o}ria a constata\c{c}\~{a}o que antipart\'{\i}culas,
diferentemente do caso com $g_{t}>1/2$, n\~{a}o s\~{a}o produzidas nas
interfaces do potencial, em outras palavras, o paradoxo de Klein n\~{a}o faz
parte do cen\'{a}rio.

\'{E} interessante observar que, quer na Classe A quer na Classe C, os
estados ligados de antipart\'{\i}culas na regi\~{a}o de intera\c{c}\~{a}o
ocorrem at\'{e} mesmo quando as part\'{\i}culas incidentes se movem com
baixas velocidades.

A aus\^{e}ncia de conflitos na Classe B, quando os resultados s\~{a}o
comparados com aqueles previstos pela mec\^{a}nica qu\^{a}ntica n\~{a}{o}%
-relativ\'{\i}stica, \'{e} facilmente explicada pela constata\c{c}\~{a}o
que, enquanto o continuum das part\'{\i}culas em $|x|>a$ ($E>mc^{2}$)
torna-se $E>mc^{2}+V_{0}$ em $|x|<a$, o continuum das antipart\'{\i}culas
\'{e} insens\'{\i}vel \`{a} a\c{c}\~{a}o da intera\c{c}\~{a}o.

\section{Estados ligados}

O formalismo desenvolvido na Se\c{c}\~{a}o 3 tamb\'{e}m permite a an\'{a}%
lise de estados ligados. Conforme j\'{a} foi mencionado, tais poss\'{\i}veis
estados ligados teriam energias no intervalo $-mc^{2}<E<mc^{2}$. As solu\c{c}%
\~{o}es para os estados ligados tamb\'{e}m podem ser segregadas em tr\^{e}s
classes, segundo o valor de $g_{t}$. \ Neste contexto, os potenciais cr\'{\i}%
ticos dados por (\ref{9f}) sofrem altera\c{c}\~{o}es de m\'{o}dulo e sinal
devido \`{a} mudan\c{c}a de $E>mc^{2}$ para $-mc^{2}<E<mc^{2}$. Temos assim
que $-2mc^{2}<V_{1}<0$, $0<V_{2}<2mc^{2}/\left( 2g_{t}-1\right) \;{\textrm{
para }}\;g_{t}>1/2$ e $-2mc^{2}/\left( 1-2g_{t}\right) <V_{2}<0\;{\textrm{
para }}\;g_{t}<1/2$. Logo a seguir exploramos algumas propriedades das
classes de solu\c{c}\~{o}es:

\medskip

\begin{itemize}
\item \textbf{Classe A - }$\mathbf{g}_{t}\,\mathbf{>1/2}$\textbf{. } Esta
classe permite estados ligados tanto para um po\c{c}o de potencial \ com $%
V_{0}<V_{1}$, quanto para uma barreira de potencial com $V_{0}>V_{2}$. O
desconforto neste \'{u}ltimo caso \'{e} aliviado pela percep\c{c}\~{a}o que
a barreira de potencial \'{e} capaz de ligar antipart\'{\i}culas. Note que
quando o po\c{c}o \'{e} extremamente raso, somente estados com $E\approx
mc^{2}$ aparecem no espectro, pois $V_{1}\rightarrow 0$ quando $E\rightarrow
mc^{2}$. Por outro lado, quando a barreira \'{e} extremamente baixa, somente
estados com $E\approx -mc^{2}$ l\'{a} surgem, pois $V_{2}\rightarrow 0$
quando $E\rightarrow -mc^{2}$. Ou seja, quando o potencial \'{e} fraco,
estados do con\-ti\-nuum de part\'{\i}culas (antipart\'{\i}culas) tornam-se
membros do espectro de estados ligados de part\'{\i}culas (antipart\'{\i}%
culas) no caso de um po\c{c}o (barreira) de potencial. Note ainda que $%
V_{1}\rightarrow -2mc^{2}$ quando $E\rightarrow -mc^{2}$, e $%
V_{2}\rightarrow 2mc^{2}/\left( 2g_{t}-1\right) $ quando $E\rightarrow
mc^{2} $, o que significa que poderiam aparecer energias pr\'{o}ximas de $%
-mc^{2}$ no espectro de part\'{\i}culas, tanto como energias pr\'{o}ximas de
$mc^{2}$ no espectro de antipart\'{\i}culas, para potenciais um tanto
intensos. \'{E} instrutivo lembrar que no caso em que $g_{t}=1$, o caso de
um potencial vetorial puro, o espectro deve exibir a simetria $E\rightarrow
-E$ sob a transforma\c{c}\~{a}o $V_{0}\rightarrow -V_{0}$.

\item \textbf{Classe B - }$\mathbf{g}_{t}\,\mathbf{=1/2}$\textbf{. } Nesta
classe, somente um po\c{c}o de potencial permite estados ligados, pois $%
V_{0}<V_{1}$. Haja vista que, tal como na classe A, $V_{1}\rightarrow 0$
quando $E\rightarrow mc^{2}$ e $V_{1}\rightarrow -2mc^{2}$ quando $%
E\rightarrow -mc^{2}$ podemos concluir que este po\c{c}o de potencial liga
somente part\'{\i}culas e que energias pr\'{o}ximas de $-mc^{2}$ s\~{a}o
toleradas para um po\c{c}o suficientemente profundo.

\item \textbf{Classe C - }$\mathbf{g}_{t}\,\mathbf{<1/2}$\textbf{. } Aqui
temos que $V_{2}<V_{0}<V_{1}$ para $E>E_{c}$ e $V_{1}<V_{0}<V_{2}$ para $%
E<E_{c}$, onde $E_{c}=-mc^{2}g_{t}/(1-gt)$. Temos que $E\approx mc^{2}$ para
os estados de part\'{\i}culas, e $E\approx -mc^{2}$ para os estados de
antipart\'{\i}culas, quando $V_{0}$ est\'{a} na vizinhan\c{c}a de seus
valores extremos. Os estados de part\'{\i}culas s\'{o} s\~{a}o consentidos
se o po\c{c}o n\~{a}o exceder a profundidade $-2mc^{2}/\left(
1-2g_{t}\right) $, enquanto a profundidade m\'{a}xima igual a $-2mc^{2}$
deve ser observada pelos estados de antipart\'{\i}culas. As energias dos
estados de antipart\'{\i}culas s\~{a}o sempre negativas. No entanto, os
estados ligados de part\'{\i}culas podem ter energias negativas se $%
-mc^{2}/\left( 1-2g_{t}\right) <V_{0}<-mc^{2}$. Em quaisquer circunst\^{a}%
ncias, os n\'{\i}veis de energias de part\'{\i}culas e antipart\'{\i}culas
nunca se interceptam. Nesta classe, conv\'{e}m lembrar, o espectro deve ser
sim\'{e}trico em torno de $E=0$ no caso de um potencial escalar puro ($%
g_{t}=0$), quando ent\~{a}o a profundidade do po\c{c}o, naturalmente, n\~{a}%
o deve ultrapassar $-2mc^{2}$ e e\-ner\-gi\-as nulas n\~{a}o s\~{a}o
permitidas. Mais ainda, h\'{a} tamb\'{e}m uma simetria do espectro em torno
de $V_{0}=-mc^{2}$, no caso de $g_{t}=0$.
\end{itemize}

A prescri\c{c}\~{a}o $k\rightarrow i\kappa $, onde
\begin{equation}
\kappa =\frac{\sqrt{m^{2}c^{4}-E^{2}}}{\hbar c}  \label{91}
\end{equation}%
com $|E|<mc^{2}$, transforma as solu\c{c}\~{o}es (\ref{5}) e (\ref{7c}) em%
\begin{equation}
\phi =\left\{
\begin{array}{c}
A_{+}\,e^{-\kappa x}+A_{-}\,e^{+\kappa x} \\
\\
C_{+}\,e^{-\kappa x}+C_{-}\,e^{+\kappa x}%
\end{array}%
\begin{array}{c}
{\textrm{para }}x<-a \\
\\
{\textrm{para }}x>+a%
\end{array}%
\right.  \label{92}
\end{equation}%
Devemos impor que $A_{+}=C_{-}=0$ para que as densidades de \textit{carga},
expressas por (\ref{rho1}), sejam finitas em $x=\pm \infty $ . Ora, tem que
ser assim, pois $\int_{-\infty }^{+\infty }dx\,|\phi |^{2}<\infty $.
Enquanto isso, a solu\c{c}\~{a}o na regi\~{a}o do potencial continua a ser
expressa por (\ref{8}) com $q$ dado por (\ref{9}). Sendo o potencial (\ref{4}%
) invariante sob invers\~{a}o espacial ($x\rightarrow -x$), podemos escolher
autofun\c{c}\~{o}es com paridades definidas, viz.%
\begin{equation}
\phi (x)=\left\{
\begin{array}{c}
A\,e^{+\kappa x} \\
\\
B\cos \left( qx\right) \\
\\
A\,e^{-\kappa x}%
\end{array}%
\begin{array}{c}
{\textrm{para }}x<-a \\
\\
{\textrm{para }}|x|<a \\
\\
{\textrm{para }}x>+a%
\end{array}%
\right.  \label{93}
\end{equation}%
para $\phi (-x)=+\phi (x)$, e

\begin{equation}
\phi (x)=\left\{
\begin{array}{c}
-A\,e^{+\kappa x} \\
\\
B\sin \left( qx\right)  \\
\\
A\,e^{-\kappa x}%
\end{array}%
\begin{array}{c}
{\textrm{para }}x<-a \\
\\
{\textrm{para }}|x|<a \\
\\
{\textrm{para }}x>+a%
\end{array}%
\right.   \label{94}
\end{equation}%
para $\phi (-x)=-\phi (x)$. Aqui, mais uma vez usamos as identidades
tri\-go\-no\-m\'{e}\-tri\-cas $\sin \left( i\theta \right) =i\sinh \left(
\theta \right) $ e $\cos \left( i\theta \right) =$ cosh$\left( \theta
\right) $ para escrever $\phi \left( x\right) $ numa forma compacta, seja $q$
uma quantidade real, seja $q$ uma quantidade imagin\'{a}ria. A continuidade
de $\phi $ e $d\phi /dx$ em $x=+a$, ou equivalentemente em $x=-a$, fornece
as condi\c{c}\~{o}es de quantiza\c{c}\~{a}o\footnote{%
Fortuitamente, as condi\c{c}\~{o}es de quantiza\c{c}\~{a}o tamb\'{e}m
poderiam ser obtidas por meio da identifica\c{c}\~{a}o dos p\'{o}los da
amplitude de transmiss\~{a}o. Para esta finalidade os valores f\'{\i}sicos
do n\'{u}mero de onda $k$, definidos no eixo real, s\~{a}o estendidos para o
plano complexo. Com efeito, o denominador de (\ref{13d}), com a prescri\c{c}%
\~{a}o $k\rightarrow i\kappa $, \'{e} nulo sempre que $\cot \left(
2qa\right) =\left( q^{2}-\kappa ^{2}\right) /\left( 2q\kappa \right) $. Com
a interven\c{c}\~{a}o da identidade $\cot \left( \theta \right) =-\tan
\left( \theta /2\right) \pm \sqrt{1+\cot ^{2}\left( \theta \right) }$,
podemos reescrever a localiza\c{c}\~{a}o dos p\'{o}los por $\tan \left(
qa\right) =\left[ \pm \left( \kappa ^{2}+q^{2}\right) +\kappa ^{2}-q^{2}%
\right] /\left( 2q\kappa \right) $, express\~{a}o que reduz-se \`{a}
primeira (segunda) linha de (\ref{95}) \ acaso o sinal \'{e} positivo
(negativo).}%
\begin{equation}
\frac{\kappa }{q}=\left\{
\begin{array}{c}
\tan \left( qa\right)  \\
\\
-\cot \left( qa\right)
\end{array}%
\begin{array}{c}
{\textrm{para }}\phi (-x)=+\phi (x) \\
\\
{\textrm{para }}\phi (-x)=-\phi (x)%
\end{array}%
\right.   \label{95}
\end{equation}

\noindent As equa\c{c}\~{o}es acima s\~{a}o equa\c{c}\~{o}es reais, seja $q$
real ou imagin\'{a}rio puro. Entretanto, as condi\c{c}\~{o}es de quantiza%
\c{c}\~{a}o para $q$ imagin\'{a}rio, viz.%
\begin{equation}
-\frac{\kappa }{|q|}=\left\{
\begin{array}{c}
\tanh \left( |q|a\right) \\
\\
\coth \left( |q|a\right)%
\end{array}%
\begin{array}{c}
{\textrm{para }}\phi (-x)=+\phi (x) \\
\\
{\textrm{para }}\phi (-x)=-\phi (x)%
\end{array}%
\right.  \label{95a}
\end{equation}

\noindent n\~{a}o fornecem solu\c{c}\~{o}es porque o membro esquerdo de (\ref%
{95a}) \'{e} negativo e os membros direitos s\~{a}o positivos. Em outras
palavras, as poss\'{\i}veis solu\c{c}\~{o}es para os estados ligados t\^{e}m
que ter um n\'{u}mero de onda real na regi\~{a}o do potencial.

Neste caso de $q\in
\mathbb{R}
$, \'{e} de utilidade definir a vari\'{a}vel $z_{0}$ como $z_{0}=\sqrt{%
z^{2}+\left( \kappa a\right) ^{2}}$, onde $z=qa$. Por conseguinte, podemos
reescrever as condi\c{c}\~{o}es de quantiza\c{c}\~{a}o como%
\begin{equation}
\sqrt{\left( \frac{z_{0}}{z}\right) ^{2}-1}=\left\{
\begin{array}{c}
\tan \left( z\right) \\
\\
-\cot \left( z\right)%
\end{array}%
\begin{array}{c}
{\textrm{para }}\phi (-x)=+\phi (x) \\
\\
{\textrm{para }}\phi (-x)=-\phi (x)%
\end{array}%
\right.  \label{96}
\end{equation}%
Note que $z<z_{0}$ por defini\c{c}\~{a}o, e que $z\simeq z_{0}$ quando $%
|E|\simeq mc^{2}$ e nos casos em que $q\gg 1/\lambda $. \ A natureza do
espectro resultante das solu\c{c}\~{o}es destas equa\c{c}\~{o}es
transcendentais podem ser visualizadas na Fig. 4, onde constam esbo\c{c}os
dos membros direito e esquerdo de (\ref{96}). As abscissas das interse\c{c}%
\~{o}es de $\tan \left( z\right) $ e $-\cot \left( z\right) $ com $\sqrt{%
\left( z_{0}/z\right) ^{2}-1}$ fornecem as solu\c{c}\~{o}es desejadas. O uso
de (\ref{9}) e (\ref{91}) permite-nos escrever%
\begin{equation}
z_{0}=\frac{a}{\hbar c}\sqrt{\left( 2g_{t}-1\right) V_{0}^{2}-2V_{0}\left[
\left( E-mc^{2}\right) g_{t}+mc^{2}\right] }  \label{z}
\end{equation}%
e da\'{\i} vemos que $z_{0}$ cresce linearmente com o aumento da largura da
regi\~{a}o de intera\c{c}\~{a}o.

Para $q\gg 1/\lambda $, o que significa $a\ll z_{0}\lambda $ e $z_{0}\simeq
z $, o que acontece somente para $g_{t}\geq 1/2$ e $|V_{0}|>2mc^{2}$, poder%
\'{a} n\~{a}o haver solu\c{c}\~{o}es de estados ligados e em algumas circunst%
\^{a}ncias a \'{u}nica solu\c{c}\~{a}o corresponde a uma autofun\c{c}\~{a}o
de paridade \'{\i}mpar.

Para $q\ll 1/\lambda $, o que significa $a\gtrsim z_{0}\lambda $ e $z_{0}\gg
z$, fica claro que h\'{a} uma seq\"{u}\^{e}ncia finita de estados ligados
com paridades alternadas. O n\'{u}mero de solu\c{c}\~{o}es cresce com o
aumento de $z_{0}$ e h\'{a} pelo menos uma solu\c{c}\~{a}o com au\-to\-fun\-%
\c{c}\~{a}o de paridade par no espectro, n\~{a}o importa o qu\~{a}o pequeno
seja $z_{0}$. Para dizer a verdade, $z_{0}$ como fun\c{c}\~{a}o de $|V_{0}|$
\'{e} monotonicamente crescente se $g_{t}\geq 1/2$, significando que o n\'{u}%
mero de estados ligados cresce com o aumento de $|V_{0}|$. Para $g_{t}<1/2$,
por\'{e}m, $z_{0}$ \'{e} crescente somente nos intervalos: $-2mc^{2}/\left(
1-2g_{t}\right) <V_{0}<-mc^{2}/\left( 1-2g_{t}\right) $ para $E>0$, e $%
-2mc^{2}<V_{0}<-mc^{2}$ para $E<0$. A Figura 4 ainda permite-nos concluir
que para grandes valores de $z_{0}$ e para os valores mais baixos de $z$, as
solu\c{c}\~{o}es de (\ref{96}) s\~{a}o expressas por $z_{n}=n\pi /2,$ onde $%
n=1,2,3,\ldots $, com $n$ \'{\i}mpar (par) correspondendo \`{a}s solu\c{c}%
\~{o}es com autofun\c{c}\~{o}es pares (\'{\i}mpares). Curiosamente, a forma
destas solu\c{c}\~{o}es assint\'{o}ticas para as energias corresponde
justamente \`{a} condi\c{c}\~{a}o de transmiss\~{a}o ressonante expressa por
(\ref{160}), sendo que agora o sinal defronte do radical deve ser escolhido
de modo que $|E|<mc^{2}$.

As Figuras 5, 6, 7, 8 e 9 ilustram os resultados do c\'{a}lculo num\'{e}rico
das solu\c{c}\~{o}es de (\ref{96}) com $V_{0}<0$ e para os menores valores
de $z_{0}$, com $g_{t}=1,\,3/4,\,1/2,\,1/4$ e $0$, respectivamente.
Consideramos $a/\lambda =1/2$ nas Figuras 5 e 6, e $a/\lambda =5$ nas
Figuras 7, 8 e 9. Tal como no caso de espalhamento, usamos o sistema de
unidades em que $\hbar =c=m=1$.

A Classe A exibe o efeito SSW para um po\c{c}o suficientemente estreito e
profundo ainda que haja um acoplamento escalar contaminante. Para po\c{c}os
largos, o comportamento do espectro n\~{a}o difere daquele resultante da equa%
\c{c}\~{a}o de Dirac, quando ent\~{a}o os n\'{\i}veis de energias
correspondentes \`{a}s part\'{\i}culas mergulham no continuum correspondente
\`{a}s antipart\'{\i}culas. Seja o po\c{c}o estreito ou largo, o paradoxo de
Klein \'{e} parte integrante do cen\'{a}rio de potenciais muito fortes. Para
a Classe B s\'{o} h\'{a} estados ligados de part\'{\i}culas e seus n\'{\i}%
veis de energia tendem assintoticamente para o con\-ti\-nuum inferior \`{a}
medida que o po\c{c}o de potencial se aprofunda, implicando assim na aus\^{e}%
ncia do paradoxo de Klein. Na Classe C, os estados ligados de antipart\'{\i}%
culas voltam a fazer parte do espectro ainda que o po\c{c}o seja pouco
profundo. Entretanto, os n\'{\i}veis de energia dos estados ligados de part%
\'{\i}culas e antipart\'{\i}culas nunca se encontram, acenando para a
completa aus\^{e}ncia do paradoxo de Klein. De qualquer jeito, quer os n%
\'{\i}veis de part\'{\i}culas, quer os n\'{\i}veis de antipart\'{\i}culas,
mergulham em seus pr\'{o}prios \textit{continua }\`{a} medida que o po\c{c}o
se torna muito profundo. H\'{a} de se notar tamb\'{e}m um fato excepcional
na Classe C: uma profundidade limite para o po\c{c}o al\'{e}m da qual n\~{a}%
o h\'{a} energias permitidas para os estados ligados, quando ent\~{a}o,
todos os n\'{\i}veis de energia j\'{a} mergulharam em seus respectivos
\textit{continua}.

O efeito SSW tem sido interpretado como sendo devido \`{a} polariza\c{c}\~{a}%
o da densidade de carga da antipart\'{\i}cula pelo potencial de intera\c{c}%
\~{a}o \cite{kle2}. Esta interpreta\c{c}\~{a}o torna-se razo\'{a}vel com um
exemplo simples da teoria eletromagn\'{e}tica cl\'{a}ssica. \'{E} sabido que
uma carga pontual pode atrair um objeto neutro em sua vizinhan\c{c}a por
causa da polariza\c{c}\~{a}o induzida pela carga puntiforme. Decerto a adi%
\c{c}\~{a}o de uma pequena quantidade de carga ao objeto, de mesmo sinal que
a carga pontual, diminuir\'{a} a for\c{c}a de atra\c{c}\~{a}o entre os dois
corpos mas n\~{a}o se pode refutar que poder\'{a} n\~{a}o ser o bastante
para causar a repuls\~{a}o. A conclus\~{a}o \'{o}bvia \'{e} que poder\'{a}
haver atra\c{c}\~{a}o entre dois corpos com cargas el\'{e}tricas de mesmo
sinal. Torna-se ainda compreens\'{\i}vel que a polariza\c{c}\~{a}o \'{e} um
fen\^{o}meno de curto alcance e requer que pelo menos um dos corpos seja
extenso, ou seja, n\~{a}o puntiforme\footnote{%
A intera\c{c}\~{a}o entre uma carga pontual e uma esfera condutora carregada
\'{e} suscet\'{\i}vel ao tratamento anal\'{\i}tico \cite{jac}, e permite
concluir que: a) para grandes dist\^{a}ncias, a for\c{c}a entre os corpos
reduz-se \`{a}quela expressa pela lei de Coulomb para duas cargas pontuais;
b) para curtas dist\^{a}ncias, h\'{a} atra\c{c}\~{a}o entre os corpos ainda
que a carga da esfera condutora tenha o mesmo sinal que a carga puntiforme.}%
. Lembrando que a \textit{carga} da part\'{\i}cula governada pela EKG est%
\'{a} distribu\'{\i}da por todo o espa\c{c}o, conforme (\ref{rho1}), podemos
inferir sobre a polariza\c{c}\~{a}o da \textit{carga} na regi\~{a}o de intera%
\c{c}\~{a}o. Com efeito, o componente vetorial do po\c{c}o de potencial faz
com que estados estacion\'{a}rios com energias negativas possam ter o sinal
de suas densidades de \textit{carga} alterados. Basta compreender que um
estado com e\-ner\-gi\-a um pouquinho maior que $-mc^{2}$ est\'{a} na imin%
\^{e}ncia de alterar o sinal de sua densidade de \textit{carga} na regi\~{a}%
o do potencial quando $|V_{0}|>mc^{2}/g_{t}$. Uma densidade de \textit{carga}
positiva na regi\~{a}o interna do po\c{c}o, n\~{a}o importando o sinal da
densidade de \textit{carga} na regi\~{a}o externa, \'{e} uma condi\c{c}\~{a}%
o \textit{sine qua non} para a exist\^{e}ncia de estados ligados.
Entretanto, torna-se imprescind\'{\i}vel que a densidade de \textit{carga}
seja tal que a intera\c{c}\~{a}o forne\c{c}a uma energia contida no conjunto
dos valores permitidos para as energias dos estados ligados. Por que o
efeito SSW n\~{a}o se manifesta quando $g_{t}\leq 1/2$? Bem, o componente
escalar do po\c{c}o de potential n\~{a}o acopla com a \textit{carga} e n\~{a}%
o interv\'{e}m na densidade de \textit{carga}. O potencial escalar age
indistintamente sobre part\'{\i}culas e antipart\'{\i}culas e assim
contribui para minimizar a efic\'{a}cia da polariza\c{c}\~{a}o devida ao
potencial vetorial. \'{E} este am\'{a}lgama de efeitos competitivos que faz
com que o efeito SSW s\'{o} se manifeste para os casos em que o coplamento
vetorial supera o acoplamento escalar.

\section{Conclus\~{a}o}

A mistura arbitr\'{a}ria de acoplamentos vetorial e escalar na EKG
unidimensional mostrou-se muito prof\'{\i}cua. Verificamos que o acoplamento
escalar n\~{a}o desempenha papel expl\'{\i}cito na determina\c{c}\~{a}o da
velocidade de grupo, e nem mesmo na determina\c{c}\~{a}o da densidade e da
corrente.

A explora\c{c}\~{a}o do potencial quadrado, tanto no caso de espalhamento
quanto no caso de estados ligados, revelou resultados realmente
surpreendentes. Em ambos os casos, a intensidade do acoplamento vetorial
relativa \`{a} intensidade do acoplamento escalar permitiu discriminar tr%
\^{e}s classes de solu\c{c}\~{o}es.

O espalhamento de part\'{\i}culas em potenciais quadrados muito intensos
apresentou re\-sul\-ta\-dos paradoxais que foram satisfatoriamente
solucionados pela suposi\c{c}\~{a}o da media\c{c}\~{a}o de antipart\'{\i}%
culas. A mistura arbitr\'{a}ria de acoplamentos desvelou a i\-ne\-xe\-q\"{u}%
i\-bi\-li\-da\-de do me\-ca\-nis\-mo da produ\c{c}\~{a}o espont\^{a}nea de
pares no caso em que $g_{t}\leq 1/2$, tanto quanto o aumento do limiar da
energia de produ\c{c}\~{a}o de pares no caso em que $g_{t}>1/2$ devido \`{a}
presen\c{c}a do acoplamento escalar. Tamb\'{e}m foi revelado que, apesar da
media\c{c}\~{a}o de antipart\'{\i}culas, somente as part\'{\i}culas s\~{a}o
irradiadas da regi\~{a}o de intera\c{c}\~{a}o.

A investiga\c{c}\~{a}o dos estados ligados em po\c{c}os de potenciais
quadrados re\-ve\-lou o surgimento do efeito SSW para potenciais intensos e
de curto alcance apenas para as circunst\^{a}ncias em que o acoplamento
vetorial excede o acoplamento escalar. Apresentamos um modelo cl\'{a}ssico
que torna plaus\'{\i}vel o efeito SSW tanto quanto sua inibi\c{c}\~{a}o pelo
potencial escalar.

Aos intr\'{e}pidos leitores, deixamos a tarefa do exame das conseq\"{u}\^{e}%
ncias da mistura de acoplamentos vetorial e escalar no potencial quadrado
para o caso fermi\^{o}nico.

\bigskip

\bigskip

\noindent \textbf{Agradecimentos:}

\medskip

Os autores s\~{a}o gratos ao CNPq e \`{a} FAPESP pelo apoio financeiro.

\newpage

\noindent \textbf{Legenda das figuras:}

\bigskip

\noindent Fig.1: Coeficiente de transmiss\~{a}o versus intensidade do potencial
para um caso da Classe A.

\bigskip

\noindent Fig.2: Coeficiente de transmiss\~{a}o versus intensidade do potencial
para um caso da Classe B.

\bigskip

\noindent Fig.3: Coeficiente de transmiss\~{a}o versus intensidade do potencial
para um caso da Classe C

\bigskip

\noindent Fig.4: Solu\c{c}\~{o}es das condi\c{c}\~{o}es de quantiza\c{c}\~{a}o
expressas por (\protect\ref{96}) para um valor representativo de $z_{0}$. A
linha cont\'{\i}nua representa a fun\c{c}\~{a}o $\protect\sqrt{\left(
z_{0}/z\right) ^{2}-1}$, a linha tracejada $\tan \left( z\right)$ e a linha
ponto-tracejada $-\cot \left( z\right)$.

\bigskip

\noindent Fig.5: N\'{\i}veis de energia em fun\c{c}\~{a}o de $V_{0}$ para para um po%
\c{c}o de potencial com $g_{t}=1$ e $a=\protect\lambda/2$.

\bigskip

\noindent Fig.6: N\'{\i}veis de energia em fun\c{c}\~{a}o de $V_{0}$ para para um po%
\c{c}o de potencial com $g_{t}=3/4$ e $a=\protect\lambda/2$.

\bigskip

\noindent Fig.7: N\'{\i}veis de energia em fun\c{c}\~{a}o de $V_{0}$ para para um po%
\c{c}o de potencial com $g_{t}=1/2$ e $a=5\protect\lambda$.

\bigskip

\noindent Fig.8: N\'{\i}veis de energia em fun\c{c}\~{a}o de $V_{0}$ para para um po%
\c{c}o de potencial com $g_{t}=1/4$ e $a=5\protect\lambda$.

\bigskip

\noindent Fig.9: N\'{\i}veis de energia em fun\c{c}\~{a}o de $V_{0}$ para para um po%
\c{c}o de potencial com $g_{t}=0$ e $a=5\protect\lambda$.

\end{document}